\documentclass[jcp,superscriptaddress,amsmath,amssymb]{revtex4}

\usepackage{graphicx}
\usepackage{dcolumn}
\usepackage{bm}

\usepackage{rotating}
\newcolumntype{.}{D{.}{.}{8}}

\usepackage[utf8]{inputenc}
\usepackage{amsmath, amssymb}
\usepackage{tabularx,booktabs,array,dcolumn}
\usepackage{setspace}
\usepackage{vmargin}
\usepackage{xcolor}
\usepackage{graphicx,psfrag,subfigure}

\usepackage{float}
\usepackage[all]{xy}

\usepackage{mathtools}
\usepackage[english]{babel}
\usepackage{bm}

\usepackage[utf8]{inputenc}
\usepackage{amsmath, amssymb}
\usepackage{tabularx,booktabs,array,dcolumn}
\usepackage{setspace}
\usepackage{vmargin}
\usepackage{graphicx,psfrag,subfigure}

\usepackage{url}
\usepackage{float}
\usepackage[all]{xy}
\usepackage{multirow}
\usepackage{comment}

\newcommand{\bos}[1]{\boldsymbol{#1}}
\newcommand{\pd}[2]{\frac{\partial #1}{\partial #2}}
\newcommand{\iim}{\mathrm{i}}
\newcommand{\cm}{cm$^{-1}$}

\def\som{Supplementary Material}

\usepackage[unicode]{hyperref}
\hypersetup{
   unicode=true,          
   plainpages=false,
   colorlinks=true,       
   citecolor=blue,        
}

\def\detg{\tilde{g}}

\def\Hrv{\hat{H}_\text{rv}}

\def\tlg{\tilde{g}}
\def\mH{m} 
\def\epsi{\varepsilon} 

\usepackage[unicode]{hyperref}
\usepackage{soul}
\hypersetup{
   unicode=true,          
   plainpages=false,
   colorlinks=true,       
   linkcolor=black,          
   linkcolor=blue,          
   citecolor=blue,        
   urlcolor=blue           
}

\begin{document}

\title{%
Rovibrational computation of H$_3^+$ with permutationally invariant Pekeris coordinates
}
\author{Gustavo Avila}
\affiliation{ELTE, E\"otv\"os Lor\'and University, Institute of Chemistry, %
P\'azm\'any P\'eter s\'et\'any 1/A, %
1117 Budapest, Hungary}

\author{Edit M\'atyus}%
\email{edit.matyus@ttk.elte.hu}
\affiliation{ELTE, E\"otv\"os Lor\'and University, Institute of Chemistry, %
P\'azm\'any P\'eter s\'et\'any 1/A, %
1117 Budapest, Hungary}
\date{\today}

\begin{abstract}
\noindent %
The Pekeris coordinates provide a permutationally invariant set of coordinates for H$_3^+$. 
They are defined as linear combinations of the three internuclear distances that automatically fulfil the triangle inequality for all non-negative coordinate values.
In this work, we test three discrete variable representations (DVR) for tightly converging the rovibrational energies up to and beyond the barrier to linearity using the Pekeris coordinates. The best performing representation is a cot-DVR-type approach adapted to the Pekeris problem. 
The two- and three-proton near coalescence region, which is also part of the direct product Pekeris grid but dynamically not relevant, is avoided by coordinate mapping and regulator functions. 
\begin{center}
  \includegraphics[width=10cm]{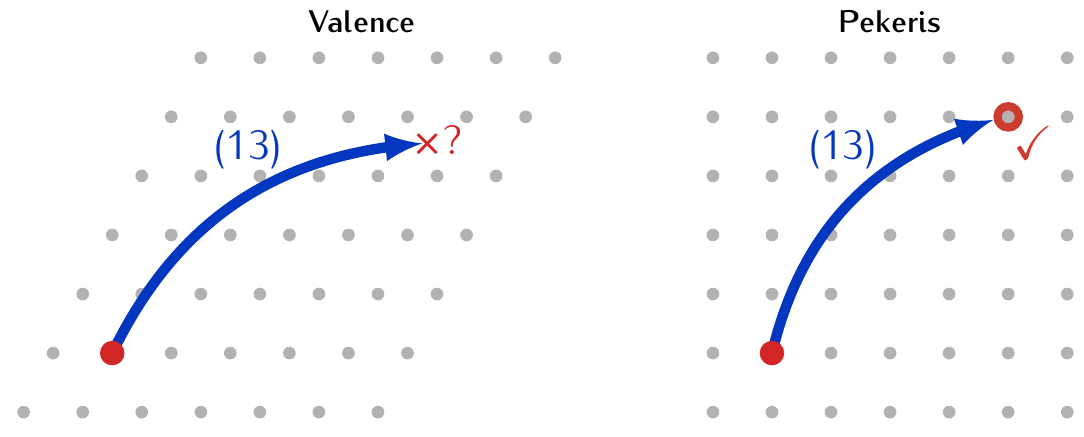}
\end{center}
%
\end{abstract}

\maketitle
\section{Introduction}
This paper is dedicated to Zlatko Ba\v{c}i{\'c} on the occasion of his birthday. Zlatko Ba\v{c}i{\'c} has been among the pioneers of rovibrational computations of floppy molecules \cite{BaLi86,BaLi87,LiBa87,BaWhBrLi88} and most recently molecular complexes in high dimensionality \cite{LiVZLiFeBa24,FeSiBa24,SiFeBa25}.

Rovibrational computations of H$_3^+$ have a long and rich history, which is briefly reviewed in the following paragraphs.
In 1976, Carney and Porter reported configuration-interaction computations for the vibrational Schrödinger equation of H$_3^+$.
In 1984, Tennyson and Sutcliffe \cite{TeSu84} used Jacobi coordinates (labelled as $R_{12}$, $R_{\text{CM}_{12}3}$ and $\theta$) in a variational approach. 
For computing the potential energy integrals, they expanded the potential energy surface (PES) in terms of Legendre functions, $V(R_{12},R_{\text{CM}_{12}3},\theta) =\sum_{\lambda=0}^{30} P_{\lambda}(\theta) V_{\lambda}(R_{12},R_{\text{CM}_{12}3})$.
In 1999, Polyansky and Tennyson \cite{PoTe99} used an \emph{ab initio} H$_3^+$ PES and effective rovibrational masses to achieve spectroscopic accuracy.

In 1994, Carrington and co-workers \cite{BrTrCaCo94} successfully combined the discrete variable representation (DVR) \cite{LiCa00} with one-dimensional tridiagonal Morse basis functions for $R_{12}$ and two-dimensional basis functions for the $(R_{\text{CM}_{12}3},\theta)$ part.
The free parameters of the basis set (three for the tridiagonal Morse basis and one in the spherical-harmonic representation) were optimised for computing the vibrational spectrum of H$_3^+$ up to dissociation. 
They developed a methodology that does not require expanding the potential energy surface and
allows one to use the most complete and accurate PES representation available. At the same time, they encountered a problem (also relevant to the present work), the occurrence of large potential energy values near coalescence points. They handled these (dynamically irrelevant) large values by introducing a potential energy cutoff, \emph{i.e.,} any potential energy larger than this cutoff was replaced by the cutoff value. This approximation limited the convergence of the vibrational energy levels to $0.01$~cm$^{-1}$ (for the actual PES). 
In Ref.~\cite{JaCa13}, Jaquet and Carrington extended these Jacobi-coordinate computations for up to $J=40$ rotational quantum numbers 
using a non-product  $K$-dependent vibrational basis set, so the basis functions vanish at the singularities of the rovibrational Hamiltonian ($K$ corresponds to the rotational angular momentum projection to the body-fixed $z$ axis).

In 1994, Hua and Carrington published a paper titled \emph{`Discrete variable representations of complicated kinetic energy operators'.} 
Among other options, they considered the special linear combination of the three internuclear distance coordinates that
could be integrated with independent intervals; in contrast to the three plain internuclear distances for which imposing the triangle inequality conditions would have been inconvenient in the DVR computations. 
The connection to Pekeris' 1958 definition of the same coordinate set~\cite{Pe58} was included during the revision stage following the observation of one of the reviewers. 
Hua and Carrington used the Pekeris coordinates with Jacobi associated functions and DVR to compute the vibrational energies of H$_{3}^{+}$. In these computations, large local PES values caused numerical problems, and instead of using a cutoff as in Ref.~\cite{BrTrCaCo94}, they discarded the three-dimensional DVR functions for which the PES was larger than 90\;000~$hc\ \text{cm}^{-1}$. 
Unlike pruning non-localised basis functions, \emph{e.g.,} Ref.~\cite{AvCa11b}, discarding localised functions limited the vibrational energy convergence to 0.1~cm$^{-1}$.

The significance of H$_3^+$ in many fields, from fundamental molecular physics, spectroscopy, through astrophysics and astrochemistry, cannot be overstated. In 2000 and in 2012, two Issues of the Philosophical Transactions of the Royal Society of London, Series A [Vol.~358, Issue~1774 and Vol.~370, Issue~1978] were dedicated to recollections about discussion meetings on fundamentals and applications of H$_3^+$ spectroscopy, \emph{e.g.,} \cite{HeMiOkWa00,Ok00,Wa00,TeKoMuPoPr00,Mc00,He00,Ok12,h3pphtrans12}.
We also mention the review by Lindsay and McCall \cite{LiMc01}, which has remained a key source and reference for H$_3^+$ spectroscopy. Most recently, a high-precision potential energy surface was developed based on explicitly correlated Gaussian computations \cite{pes12,h3pphtrans12} (which we also use in the present work).

Given this extremely rich history, can we add anything to H$_3^+$, especially in the low-energy range?
We have recently encountered the need for the explicit computation of the permutational symmetry projector of the three protons of H$_3^+$ described by the $S_3$ permutation group. We had to project rovibrational-proton-spin states onto the irreducible representation of $S_3$ that fulfils the Pauli principle, \emph{i.e.,} anti-symmetric to the exchange of any pairs of protons. 
This projection was necessary to compute the hyperfine-Zeeman splittings of the rovibrational spectrum of H$_3^+$~\cite{AvSuKoMa25}. This work also required a very high level of numerical error control, \emph{i.e.,} numerical convergence of the eigenvalues to $10^{-5}\ hc\ \text{cm}^{-1}$ (or better), within a well-defined vibrational model (PES representation and proton masses), so that the numerical noise does not blur the small magnetic shifts and splittings.

In Ref.~\cite{AvSuKoMa25}, we used valence coordinates and DVR, for which the 3-dimensional (3D) grid is not closed under all $S_3$ operations. In the present work, we explore computational approaches using the permutationally invariant Pekeris coordinates and observe that the rovibrational energy levels can be finally converged to high numerical precision for $J=0, 1$ and 2 working with a grid that is closed under all $S_3$ operations. As a result, this Pekeris representation may be a viable option for future hyperfine-Zeeman computations of H$_3^+$ in higher rovibrationally excited states.

\section{Theoretical details \label{sec:theo}}
Rovibrational computations can be efficiently performed using curvilinear, rotational and internal (vibrational) coordinates. Jacobi coordinates, Radau coordinates or simple (non-orthogonal) valence coordinates are often used. 
The general bond-angle kinetic energy operator for triatomics has been formulated by Sutcliffe and Tennyson \cite{SuTe91}. 
Furthermore, we mention the numerical kinetic energy operator (KEO) approaches~\cite{LaNa02,YuThJe07,YaYu15,MaCzCs09,FaMaCs11}. In this work, we use and further develop the numerical KEO approach implemented in the GENIUSH program~\cite{MaCzCs09,FaMaCs11} to solve the rovibrational (rv) Schrödinger equation of an $N$-atomic system with $D$ active vibrational degrees of freedom (here $N=D=3$) 
\begin{align}
  \hat{H}_\text{rv}\Psi_n &= E_n\Psi_n \; ,\nonumber \\
  \Hrv
  &=
  \frac{1}{2} 
  \sum_{k=1}^{D+3} \sum_{l=1}^{D+3} 
    \detg^{-\frac{1}{4}} \hat{p}_k G_{kl} \detg^{\frac{1}{2}} \hat{p}_l \detg^{-\frac{1}{4}} 
  + V 
  \label{eq:Hrvpod}  
  \\
  &=
  \frac{1}{2} 
  \sum_{k=1}^{D} \sum_{l=1}^{D} 
    \hat{p}_k G_{kl} \hat{p}_l 
  + 
  \frac{1}{2} 
  \sum_{k=1}^{D} \sum_{a=1}^3
  (\hat{p}_k G_{k,D+a} 
  + 
  G_{k,D+a}\hat{p}_k)\hat{J}_a \nonumber \\
  & + \frac{1}{2} \sum_{a=1}^3 G_{D+a,D+a} \hat{J}_a^2 
   + \frac{1}{2} \sum_{a=1}^3 \sum_{b>a}^3 G_{D+a,D+b}[\hat{J}_a,\hat{J}_b]_+ 
  + U + V \; .
  \label{eq:Hrv}    
\end{align}
$V$ is the potential energy surface from electronic structure theory, $\hat{p}_k=-\iim\partial / \partial q_k,\ k=1,\ldots,3N-6,$ is the generalised momentum (in atomic units) conjugate to the $q_k$ curvilinear internal (vibrational) coordinate, and
$\hat{p}_{D+a}=\hat{J}_a,\ a=1(x),2(y),3(z)$ labels the $a$th component of the rotational angular momentum in the frame fixed to the vibrating molecule (body-fixed, BF frame).
$\bos{G}=\bos{g}^{-1}$ and $\detg=\text{det}\bos{g}$ include the $\bos{g}$ mass-weighted metric tensor of the Cartesian to curvilinear coordinate transformation used to define the rotational and vibrational coordinates. $U$ is the so-called pseudopotential term, which includes coordinate derivatives of the (determinant of the) metric tensor~\cite{MaCzCs09}.
After the internal coordinates and the body-fixed frame are defined (Sec.~\ref{sec:coord}), the kinetic energy operator (KEO) coefficients are computed over the integration grid (Sec.~\ref{sec:KEO}), a basis set is chosen, and the Hamiltonian matrix is constructed by (numerical) integration (Sec.~\ref{sec:basis}). The (ro)vibrational eigenvalues and eigenvectors are computed in an iterative Lanczos procedure, \emph{e.g.,} Ref.~\citenum{MaSiCs09,SaCsAlWaMa16,SaCs16}.

\subsection{The coordinate dilemma \label{sec:coord}}
In H$_3^+$, there are three identical nuclei (protons), and the question emerges whether we can find an internal coordinate definition which is invariant under the interchange of the three nuclei. 
It would be ideal to have three internal coordinates, for which the (DVR) grid representation is closed under all permutations of the three-element permutation group, $S_3$.

\begin{figure}
  \includegraphics[scale=1.]{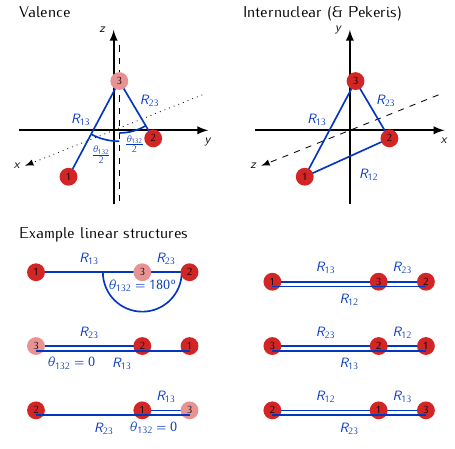}
  \caption{%
    Valence (left) and internuclear distance (right) coordinate representations. The Pekeris coordinates, Eq.~\eqref{eq:Pekeris}, are linear combinations of the internuclear distances.
    \label{fig:coord}
  }
\end{figure}

H$_3^+$ is a floppy system, the linear structures belong to the dynamically relevant range even for the lower-energy rotation-vibration motions, so it is necessary to account for the equivalent linear structures, \emph{e.g.,}
H$_{1}$-H$_{2}$-H$_{3}$, 
H$_{3}$-H$_{1}$-H$_{2}$, 
and 
H$_{2}$-H$_{3}$-H$_{1}$. 
A valence (or Jacobi or Radau) coordinate representation requires large coordinate intervals to cover all the dynamically relevant coordinate regions, and even then, we would account for the different (equivalent) structures in a non-symmetric manner, which affects the convergence of the degenerate levels. For large coordinate intervals, certain grid points may fall outside the meaningfully represented region of the PES in practical computations.

The three bond distances, $R_{12},R_{13},R_{23}\in [0,\infty)$, naturally provide a permutationally invariant representation, but their application is bound by interrelated integration intervals to fulfill the triangle inequality~\cite{Wa94,MaCzCs09}. In 1994, Watson~\cite{Wa94} used the internuclear distances to compute the rovibrational states of H$_3^+$ and assigned a large potential value (a potential wall) to those grid points which did not fulfil the triangle inequality. This numerical approach allowed him to approximately compute states, using $D_{3\text{h}}$(M) symmetry-adapted basis functions, up to the barrier to linearity.

Pekeris' \cite{Pe58} linear combination of the three internuclear distances, also called perimetric coordinates,
\begin{align}
  p_{1}=\frac{1}{2}(R_{13}+R_{12}-R_{23}), \nonumber \\
  p_{2}=\frac{1}{2}(R_{12}+R_{23}-R_{13}), \nonumber \\
  p_{3}=\frac{1}{2}(R_{13}+R_{23}-R_{12}), 
  \label{eq:Pekeris}
\end{align}
allows us to continue working with permutationally invariant coordinates, while $p_1,p_2,p_3\in [0,\infty)$ automatically fulfil the triangle inequality, so they can be numerically integrated with independent intervals.

To account for the linear H$_3^+$ nuclear configurations (Fig.~\ref{fig:coord}), which are in the dynamically important configuration range, we must have points in the 3D grid
close to the linear structures with $\mathcal{P}^\text{lin}_{123}=(p_1,p_2,p_3)=(p,p',0), (p,0,p'),$ and $(0,p,p')$ values, $p,p'\in [0,p_\text{max}]$ ($p_\text{max}$ is a maximal 1D grid value of the numerical integration). 
In a direct product quadrature grid, this also means that we have $\mathcal{P}_{123}$ points close to $(p,0,0)$, $(0,p,0)$, $(0,0,p)$, as well as $(0,0,0)$, which correspond to the two- and three-proton coalescence points, for which the PES is singular. 

\subsection{Mapping out the singular PES and KEO points \label{sec:mapping}}
In a direct product Pekeris grid, in addition to the dynamically important (near) linear structures, 
there will be near two- and three-proton (coalescence) structures, for which the PES is singular, and the neighbourhood of these singular points is not well represented by the PES function (subroutine) in practical computations.
The two- and three-proton near-coalescence points are not dynamically important in our work, and we have first attempted to eliminate their effect on the rovibrational energies by assigning some large (artificial) potential energy values to them, similarly to the potential wall idea by Hua and Carrington~\cite{HuCa94}. 

We have tested various (smooth) potential wall functions for this purpose, with a $V^\text{th}$ pre-defined, large threshold value, \emph{e.g.,}
\begin{align}
  V\;\Rightarrow\; %
  V'= V^\text{th}\text{arctg}\frac{V}{V^\text{th}} \;,
  \label{eq:Vth1}
\end{align}
or
\begin{align}
  V\;\Rightarrow\; %
  V'
  =
  \frac{V}{[1+ |\frac{V}{V^\text{th}}|^n]^{\frac{1}{n}}}, \; n\in\mathbb{N} \; ,
  \label{eq:Vth2}
\end{align}
but we could not numerically sufficiently tightly (to $10^{-4}$--$10^{-5}$~$hc~$\cm) converge the (ro)vibrational energies.
Without using a $V'$ switching function in a direct-product DVR computation, it would be impossible to (approximately) converge the rovibrational levels above the barrier to linearity due to the singularities.

\begin{figure}
  \includegraphics[width=8cm]{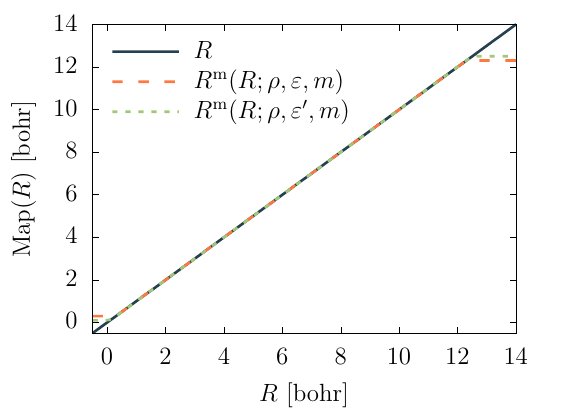}
  \caption{%
    Visualization of the effect of the coordinate mapping function, Eq.~\eqref{eq:Rmap}, for $\rho=6.30$~bohr, $\epsi=0.30$~bohr and $\epsi'=0.15$~bohr, $m=120$. 
    \label{fig:Rmap}
  }
\end{figure}

A better numerical alternative was found by exploiting a simple observation. The potential energy (and also the $U$ pseudopotential term) is a function of the internuclear distances, so we defined a mapping function for the internuclear distances, so the particles smoothly avoid the collision (coalescence) points,
\begin{align}
  R^\text{m}_n(R_n;\rho,\epsi,m)
  &=
  \frac{R_n-\rho}{[1+|\frac{R_n-\rho}{\rho-\epsi}|^m]^{\frac{1}{m}}} + \rho \; ,\quad n=1,2,3 \; .
  \label{eq:Rmap}
\end{align}
The PES (and the pseudopotential term) are calculated with these `mapped' coordinate values, 
$V(R^\text{m}_{12},R^\text{m}_{13},R^\text{m}_{23})$ as a function of the original (Pekeris) internal coordinates. As a result, the dynamically irrelevant and singular PES values are always avoided with a large potential cutoff smoothly assigned to them.
After experimenting with the $\epsi$ and $m$ parameters, we
decided to use (Sec.~\ref{sec:numres})
$m=120$, $\epsi=0.30$~bohr and $\rho=6.30$~bohr, for which $R^\text{m}$ is almost exactly identical with $R$ over the [0.5,12]~bohr interval (Fig.~\ref{fig:Rmap}). We have also tested $\epsi=0.15$ and 0.10~bohr, and the vibrational energies differ by less than $10^{-5}\ hc$~\cm\ (Fig.~\ref{fig:convJ0}).
The computed rovibrational energies remained stable to high precision to variations in the parameterisation of Eq.~\eqref{eq:Rmap}.

%
%
\subsection{Kinetic energy operator coefficients \label{sec:KEO}}
The numerical KEO approach of the GENIUSH program computes the kinetic energy operator coefficients, Eqs.~\eqref{eq:Hrvpod}--\eqref{eq:Hrv}, over a coordinate grid. 
The primary quantity is the mass-weighted metric tensor, $\bos{g}$, which is constructed with the t-vector formalism. The vibrational t-vectors are computed by finite differences of the body-fixed Cartesian coordinates with respect to the internal (here: Pekeris) coordinates \cite{MaCzCs09,MaDaAv23}. Instead of the default $10^{-5}$ (atomic unit, bohr for Pekeris coordinates) step size \cite{MaCzCs09,DaAvMa21mw}, we use the finite difference step size, $10^{-7}$, $10^{-8}$, $10^{-9}$ in atomic unit (and even smaller, with quadruple precision arithmetic) and compare the vibrational $G$ block with the analytic expressions derived by Hua and Carrington~\cite{HuCa94}
\begin{align}
  G_{11}^\text{v}
  &=
  \frac{p_1}{\mH}
  \left[%
    \frac{p_1+p_2+p_3}{R_{12} R_{13}}
    +
    \frac{p_2}{R_{13} R_{23}}
    +
    \frac{p_3}{R_{12} R_{23}}
  \right] 
  \label{eq:analG11}
  \\
  G_{22}^\text{v}
  &=
  \frac{p_2}{\mH}
  \left[%
    \frac{p_1+p_2+p_3}{R_{12} R_{23}}
    +
    \frac{p_1}{R_{13} R_{23}}
    +
    \frac{p_3}{R_{12} R_{12}}
  \right] \\     
  G_{33}^\text{v}
  &=
  \frac{p_3}{\mH}
  \left[%
    \frac{p_1+p_2+p_3}{R_{13} R_{23}}
    +
    \frac{p_1}{R_{23} R_{12}}
    +
    \frac{p_2}{R_{12} R_{13}}
  \right]  \\
  G_{12}^\text{v} &= -\frac{p_{1}p_{2}}{\mH R_{13} R_{23}} \\
  G_{13}^\text{v} &= -\frac{p_{1}p_{3}}{\mH R_{12} R_{23}}\\
  G_{23}^\text{v} &= -\frac{p_{2}p_{3}}{\mH R_{12} R_{13}} \; ,
\end{align}
and the pseudopotential term is
\begin{align}
  U (p_{1},p_{2},p_{3})
  =
  U(R_{12},R_{13},R_{23})
  =
  \frac{%
    R_{12}^4 + R_{13}^4 + R_{23}^4
  }{ 8\mH R^2_{12} R^2_{13} R^2_{23} } \; .
  \label{eq:analU}
\end{align}
The numerically computed (numerical) $G_{ij}$ and $U$ values agree with the analytic expressions in Eqs.~\eqref{eq:analG11}--\eqref{eq:analU} to better than $10^{-10}$ and $10^{-7}$ (in atomic units and $E_\text{h}$ for $U$), respectively. 
The larger deviations were observed close to the two- and three-particle coalescence points with two or three Pekeris coordinates smaller than $10^{-3}$~bohr. In these cases, the stable evaluation of even the analytic expressions required quadruple precision.

\begin{figure}
  \includegraphics[scale=0.8]{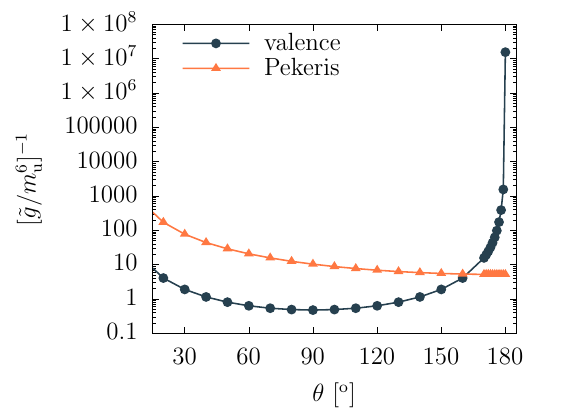}
  \caption{%
    Inverse determinant of the metric tensor in valence and Pekeris coordinates as a function of the $\theta$ valence angle, Fig.~\ref{fig:coord}.   Two proton-proton distances are fixed at $R_{13},R_{23}=1.650$~bohr. $1\ m_\text{u}=1\;822.888\;486\;3\ m_\text{e}$.
    \label{fig:invdetg}
  }
\end{figure}

In comparison with the valence coordinates (Fig.~\ref{fig:coord}), it is interesting to note that while the determinant of the inverse metric tensor, $\tlg^{-1}$ is singular at $\theta=180^\text{o}$ for valence coordinates, it is finite in Pekeris coordinates (Fig.~\ref{fig:invdetg}). 
Even though the non-singular behaviour of $\tlg^{-1}$ for Pekeris coordinates at linear structures may be an appealing property, still $\tlg$ changes by several orders of magnitude over the multi-dimensional Pekeris grid. 
For example, 
$\tlg = 1.66\cdot 10^{22}\ m_\text{e}^6=451\ m^6_\text{u}$ at the $\mathcal{P}_{123}=(12,12,10^{-4})$ near linear (and elongated) structure, 
whereas it is
$\tlg = 8.87\cdot 10^{-8}\ m_\text{e}^6=2.41\cdot 10^{-27}\ m^6_\text{u}$
at the near three-proton coalescence point, $\mathcal{P}_{123}=(10^{-4},10^{-4},10^{-4})$, which is also part of the direct product grid.

For better numerical stability, instead of the Podolsky form, Eq.~\eqref{eq:Hrvpod}, we use the rearranged form, Eq.~\eqref{eq:Hrv}. Furthermore, we replace the numerically computed $U$ with the analytic expression, Eq.~\eqref{eq:analU}. We can `regularise' $U$ with the coordinate mapping technique, Eq.~\eqref{eq:Rmap}, at near-coalescence points of the direct-product Pekeris grid.

At linear configurations, the rovibrational $\bos{G}$ tensor has singular element(s) due to a vanishing moment of inertia. In the present Pekeris computations, our body-fixed (BF) frame definition aligns the linear structures along the $z$ axis (Fig.~\ref{fig:coord}). For this BF frame, the only singular element is $G^\text{r}_{zz}$ of the rotational block (which is $G_{66}$ in Eq.~\eqref{eq:Hrv}). 
We aim to find a DVR that can accurately converge the vibrational and rovibrational energy levels using the same underlying vibrational grid. If we can find such a representation, then the transition moment integrals (\emph{e.g.,} for intensity computations or hyperfine-Zeeman simulations \cite{AvSuKoMa25}) can be computed straightforwardly in the DVR. The difficulty lies with the dynamical importance of linear structures and the $G^\text{r}_{zz}$ singularity in the rotational Hamiltonian.

\section{Vibrational basis and grid representations for (ro)vibrational computations \label{sec:basis}}
To converge rovibrational states in spite of the singular $G_{zz}^\text{r}$ element, we have tested three options. 
First, we defined a simple cutoff for $G_{zz}^\text{r}$ in Eq.~\eqref{eq:Vth1} (by replacing $V^\text{th}$) with $G_{zz}^{\text{r},\text{th}}=3.0$ (in units of $m_\text{e}^{-1}\text{bohr}^{-2}$) and used plain Laguerre-DVR, $L_n^{(0)}$.

\begin{figure}
  \includegraphics[scale=0.8]{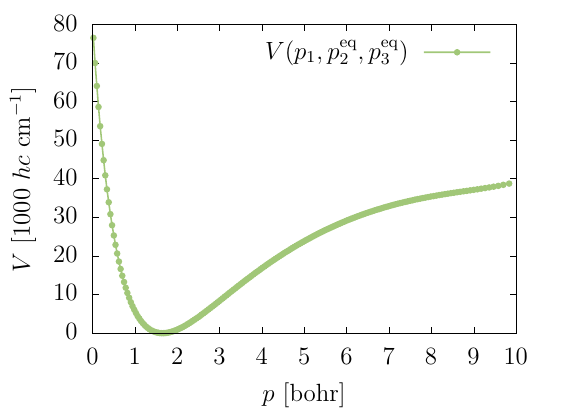}
  \caption{%
    Example effective potential energy curve used to construct the PO-DVR employed in the computation. The 1D cut of the PES \cite{pes12} is shown along one of the Pekeris coordinates, while the other two Pekeris coordinates are fixed at their equilibrium value  %
    ($p^\text{eq}=1.65$~bohr). 
    \label{fig:V1d}
  }
\end{figure}

As a second option, since the $G^\text{r}_{zz}$ singularity is repulsive (non-negative), we tried to model its effect with an appropriately chosen effective potential energy curve in a potential-optimised DVR~\cite{WeCa92,EcCl92} approach. 
We have experimented with a series of 1D model potentials; the example shown in Fig.~\ref{fig:V1d} is used in the PO-DVR (ro)vibrational computations of Sec.~\ref{sec:numres} constructed with $L_n^{(2)}$ Laguerre polynomials scaled to the $[p_\text{min},p_\text{max}]=[0,12]$~bohr interval.

In Ref.~\cite{WaCa12}, Wang and Carrington discuss techniques for handling a similar type of repulsive singularity in $K$-dependent and also $K$-independent vibrational basis sets and DVR.
Their $K$-independent approach is related to Ref.~\cite{ScMa10}, in which 
Schiffel and Manthe proposed an approach for direct product computations with repulsive singularity (singularities) of the polar angle(s) in curvilinear coordinates. In their so-called `cot-DVR' approach, instead of diagonalizing the coordinate matrix ($\cos\theta$ for the polar angle), they diagonalised $\frac{\cos\theta}{\sin\theta} = \text{cot}\theta$ (hence the name of the approach) to have a better quadrature to integrate the singular kinetic energy matrix elements. Furthermore, the finite subset of the (normalised) $\cos(n\theta)$ series of Legendre functions was appended with additional sine functions ($\sin\theta$ and $\sin2\theta$) to improve the basis convergence.

In this work, we have tried and adapted the cot-DVR idea to the 1D Pekeris problem. So, instead of the coordinate ($p$) operator, we construct the matrix
\begin{align}
  X_{ij} = \langle L^{(0)}_i |\frac{1}{p^{m}}p | L^{(0)}_j \rangle
  \label{eq:pcoord}
\end{align}
with the $L^{(0)}_n$ Laguerre functions and a series of not necessarily integer values of $m>0$, \emph{e.g.,} $m=0.5,m=0.99,m=1.5,\ldots$ up to $m=1.99$. Numerical examples are presented in Sec.~\ref{sec:numres} with $m=1.99$ (for which the regulator functions were essential,  \emph{vide infra).}
The first-order differential operator matrix was calculated as
\begin{align}
  D_{ij} = \langle L^{(0)}_i | \pd{}{p} L^{(0)}_j \rangle  \; ,
  \label{eq:pder}
\end{align}
and then, transformed using the eigenvectors of the $\bos{X}$ matrix to obtain the $\bos{D}^\text{DVR}$ matrix to construct the (ro)vibrational Hamiltonian matrix(-vector multiplication as \emph{e.g.,} in Ref.~\cite{MaCzCs09} of the implementation).
The $\bos{X}$ and $\bos{D}$ matrices, Eqs.~\eqref{eq:pcoord} and \eqref{eq:pder}, were computed by quadrature, resulting in numerically exact matrix elements. 
We note that, unlike for the $L_n^{(2)}$ associated Laguerre functions, the $\bos{D}$ matrix is not antisymmetric, so we use the matrix and also its transpose to have a symmetric Hamiltonian matrix representation for Eq.~\eqref{eq:Hrv}.
Still, the advantage of using $L^{(0)}_n$ is that it can represent vibrational wave functions with non-vanishing amplitude near linearity, \emph{i.e.,} when a single $p_i$ Pekeris coordinate approaches zero.
We note that the cot-DVR scheme~\cite{ScMa10} as well as the present p-Laguerre-DVR place more points closer to the singularity (for better integration). At the same time, the first-order derivative matrix elements are also larger, which affects the condition number of the Hamiltonian matrix \cite{DaAvMa21mw}. 

For being able to use large basis sets, we set out to better understand the origin and possibly eliminate the 
very large derivative matrix elements of DVR functions close to zero, which caused the rapidly increasing condition number of the Hamiltonian for increasing DVR sizes. 
We observed that the effect of the excessively large derivative matrix elements could be attenuated by using regulator functions for the corresponding $G_{ij}$ coefficients of the kinetic energy operator, Eq.~\eqref{eq:Hrv}. The affected $G_{ij}$ elements were at coordinate regions where the potential energy was also very large, and hence, we could discard their effect for computing the bound rovibrational states. After numerical experimentation, we regularized the vibrational block $G_{ij}\ (i,j=1,2,3)$ coefficients  as 
\begin{align}
  G_{ij}^{\text{reg}} 
  &= 
  G_{ij} F^{\text{reg}}_\text{v} \; .
  \label{eq:regGvib}
\end{align}
Various regulator functions were tested, in particular, 
\begin{align}
  F^{\text{reg}_1}_\text{v}(R_1,R_2,R_3)
  =
  \prod_{n=1}^3
    \frac{1}{1+{\rm exp}[-140(R_n/\text{bohr}-0.050)]}
  \label{eq:Fvreg1}
\end{align}
and 
\begin{align}
  F^{\text{reg}_2}_\text{v}(R_1,R_2,R_3)
  =
  \prod_{n=1}^3  
    \frac{1}{1+{\rm exp}[-140(R_n/\text{bohr}-0.025)]} \; .
  \label{eq:Fvreg2}  
\end{align}
Both of them smoothly reduce the value of the $G_{ij}^\text{v}$ coefficients to zero for
$r_{1},r_{2},r_{3}<\delta$ interatomic distances, otherwise $G_{ij}^\text{v}$ is unaffected.
The vibrational states computed with the two regulator functions are almost identical for the grid sizes used in Sec.~\ref{sec:numres}. Without these regulator functions, the Lanczos iteration remained numerically stable (in double precision arithmetic) only for smaller basis sizes.

In rovibrational computations, we need to define a regulator function also for the rotational $G^\text{r}_{ij}$ elements, which are singular due to particle-particle coalescences, 
\begin{align}
  F^\text{reg}_\text{r}(R_1,R_2,R_3)
  =
  \left[%
    \prod_{n=1}^3
    \frac{1}{1+{\rm exp}[-40(R_n/\text{bohr}-0.4)]}
  \right]^{\frac{1}{2}} \; .
  \label{eq:Frreg}
\end{align}
So, we regulated the rotational $G_{ij}$\ ($i,j>3$) elements as
\begin{align}
  G_{ij}^{\text{reg}} = G_{ij} F^{\text{reg}}_{\text{r}} 
\end{align}
and the rovibrational elements ($i=1,\ldots,3$ and $j=4(x),5(y),6(z)$ or $i,j$ exchanged) as
\begin{align}
  G_{ij}^{\text{reg}} 
  = 
  G_{ij} %
  \sqrt{F^{\text{reg}_1}_\text{v}F^{\text{reg}}_\text{r}} \; .
\end{align}

For comparison, we also show results with the already well-established and well-converged valence coordinate computations (Fig.~\ref{fig:coord}). The disadvantage of the valence coordinates was that they are not closed under all operations of the three-element permutation group. 
Regarding the Pekeris coordinates, convergence difficulties were noted in their first use in (ro)vibrational computations \cite{HuCa94}, and the rovibrational states are tightly converged in the present work (for which the application of regulator functions was essential).

Regarding the computational details for the valence coordinates, we used a potential-optimised DVR based on associated Laguerre-DVR ($L_n^{(2)}$) for $R_{1}$ and $R_{2}$. For $\cos\theta$, plain Legendre DVR was used for $J=0$ and cot-DVR (with the additional $\sin\theta$ and $\sin2\theta$ functions)  for $J>0$. For the results reported in this paper, we used 101 PO-DVR points for $R_1,R_2$ and 141 (cot-)Legendre-DVR points for $\cos\theta$ for $J=0$ ($J>0$). Regarding the convergence of these valence computations, the lowest-energy states have energies converged to $10^{-5}$~$hc$~\cm, but for some higher-energy (and high-bending states), the convergence may be less good, only $10^{-2}$~$hc$~\cm, and the Pekeris results are better converged.

\section{Numerical results \label{sec:numres}}
The rovibrational computations were carried out with the ($p_1,p_2,p_3$) Pekeris coordinates, and also for comparison with the ($R_1,R_2,\cos\theta$) valence coordinates,  using the basis and grid representations described in Sect.~\ref{sec:basis}. The potential energy surface was taken from Ref.~\cite{pes12} (GLH3P).
During the computations, we used the nuclear mass of the proton, $\mH=1.007\,276\,466\,578\,9\ m_\text{u}$.
We utilise this value in this work and note that slightly different values have already been used \cite{PoTe99} to achieve closer agreement with experiment, by modelling small non-adiabatic effects~\cite{MaTe19}.
Further conversion factors, used in this work, are $1\ m_\text{u}=1\;822.888\;486\;278\;14\ m_\text{e}$ and
$1\ E_\text{h} =219\;474.631\;363\ hc\ \text{cm}^{-1}$~\cite{codata22}. ($E_\text{h}$ is \text{hartree} and 
$1\ \text{bohr}=5.291\;772\;105\;44\  10^{-11}$~m).

\begin{figure}
  \includegraphics[scale=0.8]{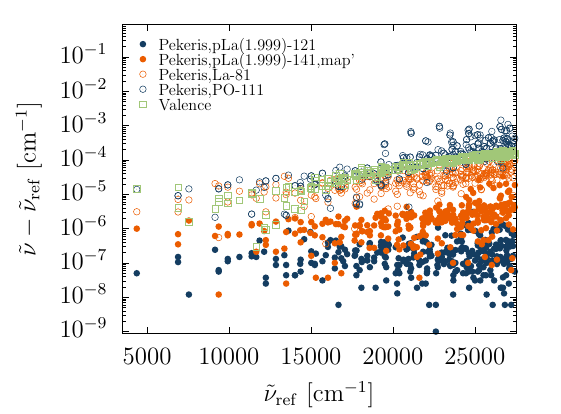}
  \caption{%
    $J=0$: %
    Convergence of the vibrational wave numbers of H$_3^+$ using Pekeris coordinates, and comparison with valence-coordinate (Valence) results is also shown. 
    For reference, we use `ref': for every $p_i$ coordinate, we use $141$ points of p-Laguerre DVR with diagonalizing the $p/p^{1.999}$ operator, $m=1.999$ in Eq.~\eqref{eq:pcoord}, $\rho=6.30$~bohr, $\epsi=0.15$~bohr, $m=120$ in the $R$ mapping, Eq.~\eqref{eq:Rmap}, and the $F_\text{v}^{\text{reg}_1}$ regulator function, Eq.~\eqref{eq:Fvreg1}.
    pLa(1.999)-121: same as `ref' but with 121 p-Laguerre($m=1.999$) points.
    pLa(1.999)-141,map': same as `ref' but with $\epsi=0.10$~bohr, in Eq.~\eqref{eq:pcoord}, and using $F_\text{v}^{\text{reg}_2}$, Eq.~\ref{eq:Fvreg2}.
    La-81: plain Laguerre DVR with 81 points.
    PO-111: PO-DVR (please see text) with 111 points.
    \label{fig:convJ0}
  }
\end{figure}

\begin{figure}
  \includegraphics[scale=0.8]{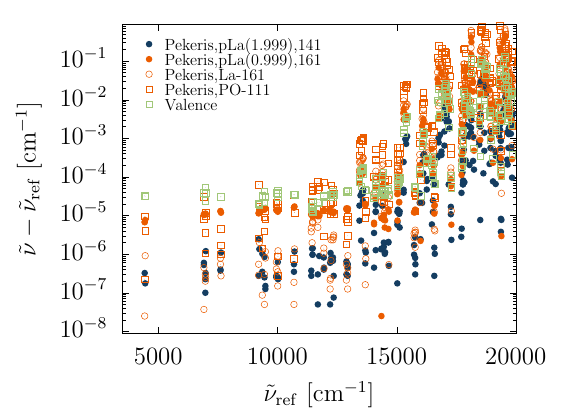}
  \caption{%
    $J=1$: %
    Convergence of the rovibrational wave numbers of H$_3^+$ using Pekeris coordinates, and comparison with valence-coordinate (Valence) results is also shown.    
    For reference, we use `ref': $161$ points of p-Laguerre DVR with diagonalizing the $p/p^{1.999}$ operator, $m=1.999$ in Eq.~\eqref{eq:pcoord}, $\rho=6.30$~bohr, $\epsi=0.15$~bohr, $m=120$ in the $R$ mapping, Eq.~\eqref{eq:Rmap}.
    pLa(1.999)-141: same as `ref' but with 141 p-Laguerre($m=1.999$) points.
    pLa(1.999)-141: same as `ref' but with p-Laguerre($m=0.999$) points.
    La-161: 161 plain Laguerre points.
    PO-111: PO-DVR (please see text) with 111 points.
    %
    \label{fig:convJ1}    
  }
\end{figure}

\begin{figure}
  \includegraphics[scale=0.8]{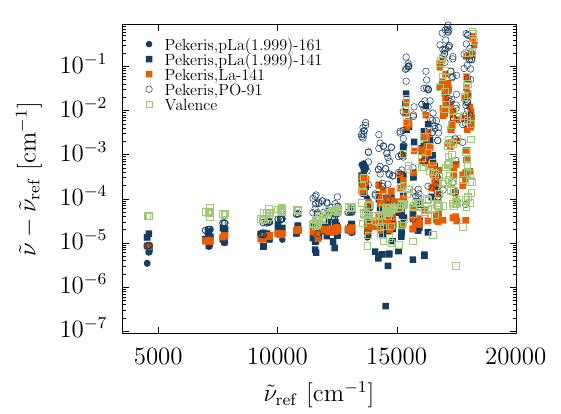}
  \caption{%
    $J=2$: %
    Convergence of the rovibrational wave numbers of H$_3^+$ using Pekeris coordinates, and comparison with valence-coordinate (Valence) results is also shown.    
    For reference, we use `ref': $141$ points of p-Laguerre DVR with diagonalizing the $p/p^{0.999}$ operator, $m=0.999$ in Eq.~\eqref{eq:pcoord}, $\rho=6.30$~bohr, $\epsi=0.15$~bohr, $m=120$ in the $R$ mapping, Eq.~\eqref{eq:Rmap}.
    pLa(1.999)-161: same as `ref' but with 161 p-Laguerre($m=1.999$) points.
    pLa(1.999)-141: same as `ref' but with p-Laguerre($m=1.999$) points.    
    La-141: 141 plain Laguerre points.    
    PO-91: PO-DVR (please see text) with 91 points.    
    \label{fig:convJ2}    
  }
\end{figure}

Figures~\ref{fig:convJ0}--\ref{fig:convJ2} show the convergence of the Pekeris computations using 
Laguerre DVR with $L^{(0)}_n$ functions (La); 
potential-optimised (PO) DVR computed with $L^{(2)}_n$ functions using $V_\text{eff}$ (Fig.~\ref{fig:V1d}); and the cot-DVR-type p-Laguerre DVR (p-La).
The convergence of the Pekeris results is tested with increasing basis sizes
and the $[p_\text{min},p_\text{max}]=[0,12]$~bohr interval for each Pekeris coordinate.
The data used to prepare Figs.~\ref{fig:convJ0}--\ref{fig:convJ2} are provided as \som.

The zero-point vibrational energy (ZPVE) is converged to $10^{-4}-10^{-5}$~$hc$~\cm\ for all cases shown in Fig.~\ref{fig:convJ0}. The vibrational energies are converged to $10^{-4}-10^{-5}$~$hc$~\cm\ in the largest computations. 
For $J=0$, plain Laguerre DVR ($k=0$) performs well, but for $J\geq 0$, the cot-DVR-type p-Laguerre representation appears to provide the best overall performance. We emphasise that the valence computations are shown for comparison and not as a benchmark; we think that the largest Pekeris computations are better converged than the valence results.
The main bottleneck of the p-Laguerre computations was the large condition number of the Hamiltonian matrix (affecting the Lanczos iteration), which was reduced by the use of the regulator functions (Sec.~\ref {sec:basis})
rendering the computations feasible in double-precision arithmetic. For the PO-DVR computations, the condition number was smaller, and regulator functions were unnecessary.

\section{Summary, conclusion, outlook}
In this work, the rovibrational energies of H$_3^+$ with $J=0,1,$ and 2 rotational quantum numbers were converged up to and beyond the barrier to linearity using Pekeris coordinates and a direct-product discrete variable representation (DVR) grid. 
Tight convergence of the energies, (by at least two-three) orders of magnitude beyond the estimated accuracy of the molecular model provided by the GLH3P PES \cite{pes12} and the nuclear masses used in this work, is necessary for a high-level of numerical error control (within the rovibrational model) to study small magnetic hyperfine-Zeeman shifts and splittings~\cite{AvSuKoMa25}. For computing rovibrational-proton-spin states, including the rotational and spin-magnetic couplings, it is required to explicitly construct the $S_3$ permutational symmetry projectors to select the Pauli-allowed states.
The Pekeris coordinates, defined as linear combinations of the internuclear distances, are closed under all permutations of the three-element permutation group, $S_3$. 
A direct-product Pekeris DVR grid is also closed under all $S_3$ operations; hence, any $S_3$ projector can be implemented using this grid representation.
Furthermore, the reported Pekeris direct-product DVR computations can be easily combined with the symmetry-adapted Lanczos scheme \cite{WaCa01}.

\vspace{0.5cm}
\begin{acknowledgments}
\noindent Financial support of the European Research Council through a Starting Grant (No.~851421) is gratefully acknowledged.
We also thank the financial support of the Hungarian National Research, Development, and Innovation Office (FK~142869), and access to the computing facility Komondor of NIIFI HPC.
\end{acknowledgments}

\clearpage

\begin{thebibliography}{45}
\expandafter\ifx\csname natexlab\endcsname\relax\def\natexlab#1{#1}\fi
\expandafter\ifx\csname bibnamefont\endcsname\relax
  \def\bibnamefont#1{#1}\fi
\expandafter\ifx\csname bibfnamefont\endcsname\relax
  \def\bibfnamefont#1{#1}\fi
\expandafter\ifx\csname citenamefont\endcsname\relax
  \def\citenamefont#1{#1}\fi
\expandafter\ifx\csname url\endcsname\relax
  \def\url#1{\texttt{#1}}\fi
\expandafter\ifx\csname urlprefix\endcsname\relax\def\urlprefix{URL }\fi
\providecommand{\bibinfo}[2]{#2}
\providecommand{\eprint}[2][]{\url{#2}}

\bibitem[{\citenamefont{Bačić and Light}(1986)}]{BaLi86}
\bibinfo{author}{\bibfnamefont{Z.}~\bibnamefont{Bačić}} \bibnamefont{and}
  \bibinfo{author}{\bibfnamefont{J.~C.} \bibnamefont{Light}},
  \bibinfo{journal}{J. Chem. Phys.} \textbf{\bibinfo{volume}{85}},
  \bibinfo{pages}{4594} (\bibinfo{year}{1986}).

\bibitem[{\citenamefont{Bačić and Light}(1987)}]{BaLi87}
\bibinfo{author}{\bibfnamefont{Z.}~\bibnamefont{Bačić}} \bibnamefont{and}
  \bibinfo{author}{\bibfnamefont{J.~C.} \bibnamefont{Light}},
  \bibinfo{journal}{J. Chem. Phys.} \textbf{\bibinfo{volume}{86}},
  \bibinfo{pages}{3065} (\bibinfo{year}{1987}).

\bibitem[{\citenamefont{Light and Bačić}(1987)}]{LiBa87}
\bibinfo{author}{\bibfnamefont{J.~C.} \bibnamefont{Light}} \bibnamefont{and}
  \bibinfo{author}{\bibfnamefont{Z.}~\bibnamefont{Bačić}},
  \bibinfo{journal}{J. Chem. Phys.} \textbf{\bibinfo{volume}{87}},
  \bibinfo{pages}{4008} (\bibinfo{year}{1987}).

\bibitem[{\citenamefont{Bačić et~al.}(1988)\citenamefont{Bačić, Whitnell,
  Brown, and Light}}]{BaWhBrLi88}
\bibinfo{author}{\bibfnamefont{Z.}~\bibnamefont{Bačić}},
  \bibinfo{author}{\bibfnamefont{R.}~\bibnamefont{Whitnell}},
  \bibinfo{author}{\bibfnamefont{D.}~\bibnamefont{Brown}}, \bibnamefont{and}
  \bibinfo{author}{\bibfnamefont{J.}~\bibnamefont{Light}},
  \bibinfo{journal}{Comp. Phys. Comm.} \textbf{\bibinfo{volume}{51}},
  \bibinfo{pages}{35} (\bibinfo{year}{1988}), ISSN \bibinfo{issn}{0010-4655}.

\bibitem[{\citenamefont{Li et~al.}(2024)\citenamefont{Li, Vindel-Zandbergen,
  Li, Felker, and Bačić}}]{LiVZLiFeBa24}
\bibinfo{author}{\bibfnamefont{J.}~\bibnamefont{Li}},
  \bibinfo{author}{\bibfnamefont{P.}~\bibnamefont{Vindel-Zandbergen}},
  \bibinfo{author}{\bibfnamefont{J.}~\bibnamefont{Li}},
  \bibinfo{author}{\bibfnamefont{P.~M.} \bibnamefont{Felker}},
  \bibnamefont{and} \bibinfo{author}{\bibfnamefont{Z.}~\bibnamefont{Bačić}},
  \bibinfo{journal}{J. Phys. Chem. A} \textbf{\bibinfo{volume}{128}},
  \bibinfo{pages}{9707} (\bibinfo{year}{2024}).

\bibitem[{\citenamefont{Felker et~al.}(2024)\citenamefont{Felker, Simkó, and
  Bačić}}]{FeSiBa24}
\bibinfo{author}{\bibfnamefont{P.~M.} \bibnamefont{Felker}},
  \bibinfo{author}{\bibfnamefont{I.}~\bibnamefont{Simkó}}, \bibnamefont{and}
  \bibinfo{author}{\bibfnamefont{Z.}~\bibnamefont{Bačić}},
  \bibinfo{journal}{J. Phys. Chem. A} \textbf{\bibinfo{volume}{128}},
  \bibinfo{pages}{8170} (\bibinfo{year}{2024}).

\bibitem[{\citenamefont{Simkó et~al.}(2025)\citenamefont{Simkó, Felker, and
  Bačić}}]{SiFeBa25}
\bibinfo{author}{\bibfnamefont{I.}~\bibnamefont{Simkó}},
  \bibinfo{author}{\bibfnamefont{P.~M.} \bibnamefont{Felker}},
  \bibnamefont{and} \bibinfo{author}{\bibfnamefont{Z.}~\bibnamefont{Bačić}},
  \bibinfo{journal}{J. Chem. Phys.} \textbf{\bibinfo{volume}{162}},
  \bibinfo{pages}{034301} (\bibinfo{year}{2025}), ISSN
  \bibinfo{issn}{0021-9606}.

\bibitem[{\citenamefont{Tennyson and Sutcliffe}(1984)}]{TeSu84}
\bibinfo{author}{\bibfnamefont{J.}~\bibnamefont{Tennyson}} \bibnamefont{and}
  \bibinfo{author}{\bibfnamefont{B.~T.} \bibnamefont{Sutcliffe}},
  \bibinfo{journal}{Mol. Phys.} \textbf{\bibinfo{volume}{51}},
  \bibinfo{pages}{887} (\bibinfo{year}{1984}).

\bibitem[{\citenamefont{Polyansky and Tennyson}(1999)}]{PoTe99}
\bibinfo{author}{\bibfnamefont{O.~L.} \bibnamefont{Polyansky}}
  \bibnamefont{and} \bibinfo{author}{\bibfnamefont{J.}~\bibnamefont{Tennyson}},
  \bibinfo{journal}{J. Chem. Phys.} \textbf{\bibinfo{volume}{110}},
  \bibinfo{pages}{5056} (\bibinfo{year}{1999}).

\bibitem[{\citenamefont{Bramley et~al.}(1994)\citenamefont{Bramley, Tromp,
  Carrington, and Corey}}]{BrTrCaCo94}
\bibinfo{author}{\bibfnamefont{M.~J.} \bibnamefont{Bramley}},
  \bibinfo{author}{\bibfnamefont{J.~W.} \bibnamefont{Tromp}},
  \bibinfo{author}{\bibfnamefont{J.}~\bibnamefont{Carrington},
  \bibfnamefont{Tucker}}, \bibnamefont{and}
  \bibinfo{author}{\bibfnamefont{G.~C.} \bibnamefont{Corey}},
  \bibinfo{journal}{J. Chem. Phys.} \textbf{\bibinfo{volume}{100}},
  \bibinfo{pages}{6175} (\bibinfo{year}{1994}), ISSN \bibinfo{issn}{0021-9606}.

\bibitem[{\citenamefont{Light and Carrington~Jr.}(2000)}]{LiCa00}
\bibinfo{author}{\bibfnamefont{J.~C.} \bibnamefont{Light}} \bibnamefont{and}
  \bibinfo{author}{\bibfnamefont{T.}~\bibnamefont{Carrington~Jr.}},
  \emph{\bibinfo{title}{Discrete-Variable Representations and their
  Utilization}} (\bibinfo{publisher}{John Wiley \& Sons, Ltd},
  \bibinfo{year}{2000}), pp. \bibinfo{pages}{263--310}, ISBN
  \bibinfo{isbn}{9780470141731}.

\bibitem[{\citenamefont{Jaquet and Carrington}(2013)}]{JaCa13}
\bibinfo{author}{\bibfnamefont{R.}~\bibnamefont{Jaquet}} \bibnamefont{and}
  \bibinfo{author}{\bibfnamefont{T.~J.} \bibnamefont{Carrington}},
  \bibinfo{journal}{J. Phys. Chem. A} \textbf{\bibinfo{volume}{117}},
  \bibinfo{pages}{9493} (\bibinfo{year}{2013}).

\bibitem[{\citenamefont{Pekeris}(1958)}]{Pe58}
\bibinfo{author}{\bibfnamefont{C.~L.} \bibnamefont{Pekeris}},
  \bibinfo{journal}{Phys. Rev.} \textbf{\bibinfo{volume}{112}},
  \bibinfo{pages}{1649} (\bibinfo{year}{1958}).

\bibitem[{\citenamefont{Avila and T.~Carrington}(2011)}]{AvCa11b}
\bibinfo{author}{\bibfnamefont{G.}~\bibnamefont{Avila}} \bibnamefont{and}
  \bibinfo{author}{\bibfnamefont{J.}~\bibnamefont{T.~Carrington}},
  \bibinfo{journal}{J. Chem. Phys.} \textbf{\bibinfo{volume}{134}},
  \bibinfo{pages}{064101} (\bibinfo{year}{2011}).

\bibitem[{\citenamefont{Herbst et~al.}(2000)\citenamefont{Herbst, Miller, Oka,
  and Watson}}]{HeMiOkWa00}
\bibinfo{author}{\bibfnamefont{E.}~\bibnamefont{Herbst}},
  \bibinfo{author}{\bibfnamefont{S.}~\bibnamefont{Miller}},
  \bibinfo{author}{\bibfnamefont{T.}~\bibnamefont{Oka}}, \bibnamefont{and}
  \bibinfo{author}{\bibfnamefont{J.~K.} \bibnamefont{Watson}},
  \bibinfo{journal}{Phil. Trans. R. Soc. A} \textbf{\bibinfo{volume}{358}},
  \bibinfo{pages}{2371} (\bibinfo{year}{2000}).

\bibitem[{\citenamefont{Oka}(2000)}]{Ok00}
\bibinfo{author}{\bibfnamefont{T.}~\bibnamefont{Oka}}, \bibinfo{journal}{Phil.
  Trans. R. Soc. A} \textbf{\bibinfo{volume}{358}}, \bibinfo{pages}{2363}
  (\bibinfo{year}{2000}).

\bibitem[{\citenamefont{Watson}(2000)}]{Wa00}
\bibinfo{author}{\bibfnamefont{J.~K.~G.} \bibnamefont{Watson}},
  \bibinfo{journal}{Phil. Trans. R. Soc. A} \textbf{\bibinfo{volume}{358}},
  \bibinfo{pages}{2371} (\bibinfo{year}{2000}).

\bibitem[{\citenamefont{Tennyson et~al.}(2000)\citenamefont{Tennyson, Kostin,
  Mussa, Polyansky, and Prosmiti}}]{TeKoMuPoPr00}
\bibinfo{author}{\bibfnamefont{J.}~\bibnamefont{Tennyson}},
  \bibinfo{author}{\bibfnamefont{M.~A.} \bibnamefont{Kostin}},
  \bibinfo{author}{\bibfnamefont{H.~Y.} \bibnamefont{Mussa}},
  \bibinfo{author}{\bibfnamefont{O.~L.} \bibnamefont{Polyansky}},
  \bibnamefont{and} \bibinfo{author}{\bibfnamefont{R.}~\bibnamefont{Prosmiti}},
  \bibinfo{journal}{Phil. Trans. R. Soc. A} \textbf{\bibinfo{volume}{358}},
  \bibinfo{pages}{2419} (\bibinfo{year}{2000}).

\bibitem[{\citenamefont{McCall}(2000)}]{Mc00}
\bibinfo{author}{\bibfnamefont{B.~J.} \bibnamefont{McCall}},
  \bibinfo{journal}{Phil. Trans. R. Soc. A} \textbf{\bibinfo{volume}{358}},
  \bibinfo{pages}{2385} (\bibinfo{year}{2000}).

\bibitem[{\citenamefont{Herbst}(2000)}]{He00}
\bibinfo{author}{\bibfnamefont{E.}~\bibnamefont{Herbst}},
  \bibinfo{journal}{Phil. Trans. R. Soc. A} \textbf{\bibinfo{volume}{358}},
  \bibinfo{pages}{2523} (\bibinfo{year}{2000}).

\bibitem[{\citenamefont{Oka}(2012)}]{Ok12}
\bibinfo{author}{\bibfnamefont{T.}~\bibnamefont{Oka}}, \bibinfo{journal}{Phil.
  Trans. R. Soc. A} \textbf{\bibinfo{volume}{370}}, \bibinfo{pages}{4991}
  (\bibinfo{year}{2012}).

\bibitem[{\citenamefont{Polyansky et~al.}(2012)\citenamefont{Polyansky, Alijah,
  Zobov, Mizus, Ovsyannikov, Tennyson, Lodi, Szidarovszky, and
  Császár}}]{h3pphtrans12}
\bibinfo{author}{\bibfnamefont{O.~L.} \bibnamefont{Polyansky}},
  \bibinfo{author}{\bibfnamefont{A.}~\bibnamefont{Alijah}},
  \bibinfo{author}{\bibfnamefont{N.~F.} \bibnamefont{Zobov}},
  \bibinfo{author}{\bibfnamefont{I.~I.} \bibnamefont{Mizus}},
  \bibinfo{author}{\bibfnamefont{R.~I.} \bibnamefont{Ovsyannikov}},
  \bibinfo{author}{\bibfnamefont{J.}~\bibnamefont{Tennyson}},
  \bibinfo{author}{\bibfnamefont{L.}~\bibnamefont{Lodi}},
  \bibinfo{author}{\bibfnamefont{T.}~\bibnamefont{Szidarovszky}},
  \bibnamefont{and} \bibinfo{author}{\bibfnamefont{A.~G.}
  \bibnamefont{Császár}}, \bibinfo{journal}{Phil. Trans. R. Soc. A}
  \textbf{\bibinfo{volume}{370}}, \bibinfo{pages}{5014} (\bibinfo{year}{2012}).

\bibitem[{\citenamefont{Lindsay and McCall}(2001)}]{LiMc01}
\bibinfo{author}{\bibfnamefont{M.}~\bibnamefont{Lindsay}} \bibnamefont{and}
  \bibinfo{author}{\bibfnamefont{B.~J.} \bibnamefont{McCall}},
  \bibinfo{journal}{J. Mol. Spectrosc.} \textbf{\bibinfo{volume}{210}},
  \bibinfo{pages}{60} (\bibinfo{year}{2001}).

\bibitem[{\citenamefont{Pavanello et~al.}(2012)\citenamefont{Pavanello,
  Adamowicz, Alijah, Zobov, Mizus, Polyansky, Tennyson, Szidarovszky, and
  Császár}}]{pes12}
\bibinfo{author}{\bibfnamefont{M.}~\bibnamefont{Pavanello}},
  \bibinfo{author}{\bibfnamefont{L.}~\bibnamefont{Adamowicz}},
  \bibinfo{author}{\bibfnamefont{A.}~\bibnamefont{Alijah}},
  \bibinfo{author}{\bibfnamefont{N.~F.} \bibnamefont{Zobov}},
  \bibinfo{author}{\bibfnamefont{I.~I.} \bibnamefont{Mizus}},
  \bibinfo{author}{\bibfnamefont{O.~L.} \bibnamefont{Polyansky}},
  \bibinfo{author}{\bibfnamefont{J.}~\bibnamefont{Tennyson}},
  \bibinfo{author}{\bibfnamefont{T.}~\bibnamefont{Szidarovszky}},
  \bibnamefont{and} \bibinfo{author}{\bibfnamefont{A.~G.}
  \bibnamefont{Császár}}, \bibinfo{journal}{J. Chem. Phys.}
  \textbf{\bibinfo{volume}{136}}, \bibinfo{pages}{184303}
  (\bibinfo{year}{2012}).

\bibitem[{\citenamefont{Avila et~al.}(2025)\citenamefont{Avila, Sunaga,
  Komorovsky, and M\'atyus}}]{AvSuKoMa25}
\bibinfo{author}{\bibfnamefont{G.}~\bibnamefont{Avila}},
  \bibinfo{author}{\bibfnamefont{A.}~\bibnamefont{Sunaga}},
  \bibinfo{author}{\bibfnamefont{S.}~\bibnamefont{Komorovsky}},
  \bibnamefont{and} \bibinfo{author}{\bibfnamefont{E.}~\bibnamefont{M\'atyus}},
  \bibinfo{journal}{Phys. Rev. Lett.}  (\bibinfo{year}{2025}).

\bibitem[{\citenamefont{Sutcliffe and Tennyson}(1991)}]{SuTe91}
\bibinfo{author}{\bibfnamefont{B.~T.} \bibnamefont{Sutcliffe}}
  \bibnamefont{and} \bibinfo{author}{\bibfnamefont{J.}~\bibnamefont{Tennyson}},
  \bibinfo{journal}{Int. J. Quant. Chem.} \textbf{\bibinfo{volume}{39}},
  \bibinfo{pages}{183} (\bibinfo{year}{1991}).

\bibitem[{\citenamefont{Lauvergnat and Nauts}(2002)}]{LaNa02}
\bibinfo{author}{\bibfnamefont{D.}~\bibnamefont{Lauvergnat}} \bibnamefont{and}
  \bibinfo{author}{\bibfnamefont{A.}~\bibnamefont{Nauts}}, \bibinfo{journal}{J.
  Chem. Phys.} \textbf{\bibinfo{volume}{116}}, \bibinfo{pages}{8560}
  (\bibinfo{year}{2002}).

\bibitem[{\citenamefont{Yurchenko et~al.}(2007)\citenamefont{Yurchenko, Thiel,
  and Jensen}}]{YuThJe07}
\bibinfo{author}{\bibfnamefont{S.~N.} \bibnamefont{Yurchenko}},
  \bibinfo{author}{\bibfnamefont{W.}~\bibnamefont{Thiel}}, \bibnamefont{and}
  \bibinfo{author}{\bibfnamefont{P.}~\bibnamefont{Jensen}},
  \bibinfo{journal}{J. Mol. Spectrosc.} \textbf{\bibinfo{volume}{245}},
  \bibinfo{pages}{126} (\bibinfo{year}{2007}).

\bibitem[{\citenamefont{Yachmenev and Yurchenko}(2015)}]{YaYu15}
\bibinfo{author}{\bibfnamefont{A.}~\bibnamefont{Yachmenev}} \bibnamefont{and}
  \bibinfo{author}{\bibfnamefont{S.~N.} \bibnamefont{Yurchenko}},
  \bibinfo{journal}{J. Chem. Phys.} \textbf{\bibinfo{volume}{143}},
  \bibinfo{pages}{014105} (\bibinfo{year}{2015}).

\bibitem[{\citenamefont{M\'atyus et~al.}(2009)\citenamefont{M\'atyus, Czak\'o,
  and Cs\'asz\'ar}}]{MaCzCs09}
\bibinfo{author}{\bibfnamefont{E.}~\bibnamefont{M\'atyus}},
  \bibinfo{author}{\bibfnamefont{G.}~\bibnamefont{Czak\'o}}, \bibnamefont{and}
  \bibinfo{author}{\bibfnamefont{A.~G.} \bibnamefont{Cs\'asz\'ar}},
  \bibinfo{journal}{J. Chem. Phys.} \textbf{\bibinfo{volume}{130}},
  \bibinfo{pages}{134112} (\bibinfo{year}{2009}).

\bibitem[{\citenamefont{F\'abri et~al.}(2011)\citenamefont{F\'abri, M\'atyus,
  and Cs\'asz\'ar}}]{FaMaCs11}
\bibinfo{author}{\bibfnamefont{C.}~\bibnamefont{F\'abri}},
  \bibinfo{author}{\bibfnamefont{E.}~\bibnamefont{M\'atyus}}, \bibnamefont{and}
  \bibinfo{author}{\bibfnamefont{A.~G.} \bibnamefont{Cs\'asz\'ar}},
  \bibinfo{journal}{J. Chem. Phys.} \textbf{\bibinfo{volume}{134}},
  \bibinfo{pages}{074105} (\bibinfo{year}{2011}).

\bibitem[{\citenamefont{Mátyus et~al.}(2009)\citenamefont{Mátyus, Šimunek,
  and Császár}}]{MaSiCs09}
\bibinfo{author}{\bibfnamefont{E.}~\bibnamefont{Mátyus}},
  \bibinfo{author}{\bibfnamefont{J.}~\bibnamefont{Šimunek}}, \bibnamefont{and}
  \bibinfo{author}{\bibfnamefont{A.~G.} \bibnamefont{Császár}},
  \bibinfo{journal}{J. Chem. Phys.} \textbf{\bibinfo{volume}{131}},
  \bibinfo{pages}{074106} (\bibinfo{year}{2009}).

\bibitem[{\citenamefont{Sarka et~al.}(2016)\citenamefont{Sarka, Császár,
  Althorpe, Wales, and Mátyus}}]{SaCsAlWaMa16}
\bibinfo{author}{\bibfnamefont{J.}~\bibnamefont{Sarka}},
  \bibinfo{author}{\bibfnamefont{A.~G.} \bibnamefont{Császár}},
  \bibinfo{author}{\bibfnamefont{S.~C.} \bibnamefont{Althorpe}},
  \bibinfo{author}{\bibfnamefont{D.~J.} \bibnamefont{Wales}}, \bibnamefont{and}
  \bibinfo{author}{\bibfnamefont{E.}~\bibnamefont{Mátyus}},
  \bibinfo{journal}{Phys. Chem. Chem. Phys.} \textbf{\bibinfo{volume}{18}},
  \bibinfo{pages}{22816} (\bibinfo{year}{2016}).

\bibitem[{\citenamefont{Sarka and Császár}(2016)}]{SaCs16}
\bibinfo{author}{\bibfnamefont{J.}~\bibnamefont{Sarka}} \bibnamefont{and}
  \bibinfo{author}{\bibfnamefont{A.~G.} \bibnamefont{Császár}},
  \bibinfo{journal}{J. Chem. Phys.} \textbf{\bibinfo{volume}{144}},
  \bibinfo{pages}{154309} (\bibinfo{year}{2016}).

\bibitem[{\citenamefont{Watson}(1994)}]{Wa94}
\bibinfo{author}{\bibfnamefont{J.~K.~G.} \bibnamefont{Watson}},
  \bibinfo{journal}{Can. J. Phys.} \textbf{\bibinfo{volume}{72}},
  \bibinfo{pages}{238} (\bibinfo{year}{1994}).

\bibitem[{\citenamefont{Wei and Carrington}(1994)}]{HuCa94}
\bibinfo{author}{\bibfnamefont{H.}~\bibnamefont{Wei}} \bibnamefont{and}
  \bibinfo{author}{\bibfnamefont{T.}~\bibnamefont{Carrington}},
  \bibinfo{journal}{J. Chem. Phys.} \textbf{\bibinfo{volume}{101}},
  \bibinfo{pages}{1343} (\bibinfo{year}{1994}).

\bibitem[{\citenamefont{Mátyus et~al.}(2023)\citenamefont{Mátyus, Martín
  Santa~Daría, and Avila}}]{MaDaAv23}
\bibinfo{author}{\bibfnamefont{E.}~\bibnamefont{Mátyus}},
  \bibinfo{author}{\bibfnamefont{A.}~\bibnamefont{Martín Santa~Daría}},
  \bibnamefont{and} \bibinfo{author}{\bibfnamefont{G.}~\bibnamefont{Avila}},
  \bibinfo{journal}{Chem. Commun.} \textbf{\bibinfo{volume}{59}},
  \bibinfo{pages}{366} (\bibinfo{year}{2023}).

\bibitem[{\citenamefont{Martín Santa~Daría et~al.}(2021)\citenamefont{Martín
  Santa~Daría, Avila, and Mátyus}}]{DaAvMa21mw}
\bibinfo{author}{\bibfnamefont{A.}~\bibnamefont{Martín Santa~Daría}},
  \bibinfo{author}{\bibfnamefont{G.}~\bibnamefont{Avila}}, \bibnamefont{and}
  \bibinfo{author}{\bibfnamefont{E.}~\bibnamefont{Mátyus}},
  \bibinfo{journal}{J. Chem. Phys.} \textbf{\bibinfo{volume}{154}},
  \bibinfo{pages}{224302} (\bibinfo{year}{2021}).

\bibitem[{\citenamefont{Wei and Carrington}(1992)}]{WeCa92}
\bibinfo{author}{\bibfnamefont{H.}~\bibnamefont{Wei}} \bibnamefont{and}
  \bibinfo{author}{\bibfnamefont{J.}~\bibnamefont{Carrington},
  \bibfnamefont{Tucker}}, \bibinfo{journal}{J. Chem. Phys.}
  \textbf{\bibinfo{volume}{97}}, \bibinfo{pages}{3029} (\bibinfo{year}{1992}),
  ISSN \bibinfo{issn}{0021-9606}.

\bibitem[{\citenamefont{Echave and Clary}(1992)}]{EcCl92}
\bibinfo{author}{\bibfnamefont{J.}~\bibnamefont{Echave}} \bibnamefont{and}
  \bibinfo{author}{\bibfnamefont{D.~C.} \bibnamefont{Clary}},
  \bibinfo{journal}{Chem. Phys. Lett.} \textbf{\bibinfo{volume}{190}},
  \bibinfo{pages}{225} (\bibinfo{year}{1992}), ISSN \bibinfo{issn}{0009-2614}.

\bibitem[{\citenamefont{Wang and {Carrington, Jr.}}(2012)}]{WaCa12}
\bibinfo{author}{\bibfnamefont{X.-G.} \bibnamefont{Wang}} \bibnamefont{and}
  \bibinfo{author}{\bibfnamefont{T.}~\bibnamefont{{Carrington, Jr.}}},
  \bibinfo{journal}{Mol. Phys.} \textbf{\bibinfo{volume}{110}},
  \bibinfo{pages}{825} (\bibinfo{year}{2012}).

\bibitem[{\citenamefont{Schiffel and Manthe}(2010)}]{ScMa10}
\bibinfo{author}{\bibfnamefont{G.}~\bibnamefont{Schiffel}} \bibnamefont{and}
  \bibinfo{author}{\bibfnamefont{U.}~\bibnamefont{Manthe}},
  \bibinfo{journal}{Chem. Phys.} \textbf{\bibinfo{volume}{374}},
  \bibinfo{pages}{118} (\bibinfo{year}{2010}).

\bibitem[{\citenamefont{M\'atyus and Teufel}(2019)}]{MaTe19}
\bibinfo{author}{\bibfnamefont{E.}~\bibnamefont{M\'atyus}} \bibnamefont{and}
  \bibinfo{author}{\bibfnamefont{S.}~\bibnamefont{Teufel}},
  \bibinfo{journal}{J. Chem. Phys.} \textbf{\bibinfo{volume}{151}},
  \bibinfo{pages}{014113} (\bibinfo{year}{2019}).

\bibitem[{cod()}]{codata22}
\bibinfo{note}{2022 {CODATA} recommended values. Fundamental constants and
  conversion factors, https://physics.nist.gov/cuu/Constants/index.html. Last
  accessed on 23 April 2025.}

\bibitem[{\citenamefont{Wang and {Carrington, Jr.}}(2001)}]{WaCa01}
\bibinfo{author}{\bibfnamefont{X.-G.} \bibnamefont{Wang}} \bibnamefont{and}
  \bibinfo{author}{\bibfnamefont{T.}~\bibnamefont{{Carrington, Jr.}}},
  \bibinfo{journal}{J. Chem. Phys.} \textbf{\bibinfo{volume}{114}},
  \bibinfo{pages}{1473} (\bibinfo{year}{2001}), ISSN \bibinfo{issn}{0021-9606}.

\end{thebibliography}

\end{document}